\documentclass{lncs/llncs}

\usepackage{graphicx}
\usepackage{tikz}

\def\pprw{8.5in}
\def\pprh{11in}

\usepackage[pdftex]{hyperref}
\hypersetup{
bookmarksnumbered,
colorlinks,
citecolor=black,
filecolor=black,
linkcolor=black,
urlcolor=black,
breaklinks=true,
}

\begin{document}

\title{The iCrawl Wizard -- Supporting Interactive Focused Crawl Specification}

\author{Gerhard Gossen \and Elena Demidova \and Thomas Risse}
\institute{L3S Research Center and Leibniz University of Hanover, Germany\\
\email{\{gossen, demidova, risse\}@L3S.de}\\
}

\maketitle
\begin{abstract}
Collections of Web documents about specific topics are needed for many areas of current research.
Focused crawling enables the creation of such collections on demand.
Current focused crawlers require the user to manually specify starting points for the crawl (\emph{seed URLs}).
These are also used to describe the expected topic of the collection.
The choice of seed URLs influences the quality of the resulting collection and requires a lot of expertise.
In this demonstration we present the iCrawl Wizard, a tool that assists users in defining focused crawls efficiently and semi-automatically.
Our tool uses major search engines and Social Media APIs as well as information extraction techniques to find seed URLs and a semantic description of the crawl intent.
Using the iCrawl Wizard even non-expert users can create semantic specifications for focused crawlers
interactively and efficiently.
\end{abstract}

\section{Introduction}
Focused crawlers \cite{chakrabarti99:focus,Pereira14} enable the efficient 
creation of topically and temporally coherent 
document collections from the Web and Social Media. Such collections are increasingly used in many domains such as digital sociology, history, politics, and journalism \cite{demidova2014arcomem,risse2014webobserv}. 
By using focused crawlers,  researchers, archivists and journalists can create sub-collections 
about specific events and topics such as the Ebola outbreak or the Ukraine crisis efficiently on demand.

Focused crawling starts with the \emph{manual} definition of the \emph{crawl specification}, 
a list of so-called \emph{seed URLs} and (optionally) keywords and entities representing 
the crawl intent of the user. 
The crawl specification is necessary for the focused crawler to efficiently 
find relevant pages and to correctly judge their relevance.
Firstly, the crawler uses seed URLs as starting points for the traversal of the Web graph, such that good 
seeds can lead the crawler directly to relevant pages. Secondly,
the content of the pages specified by the seed URLs
is used to perform relevance estimation of unseen documents collected during the crawling.
Thus, the success of the focused crawlers depends on the expertise of the user to specify representative seed URLs and keywords.
However, our anticipated non-expert users need to create collections only rarely and thus cannot develop the necessary experience.

In this demonstration we present the iCrawl Wizard\footnote{The demo is available online 
at \url{http://icrawl.l3s.uni-hannover.de:8090/campaign/1/add}}, novel interface that enables non-expert users to interactively and efficiently create 
the crawl specification for a focused crawler. 
The iCrawl Wizard combines Web search, Social Media queries and information extraction tools 
to enable users to compose crawl specification efficiently in an intuitive way.
It builds upon users' previous experience with Web search engines and allows them to start the  crawl specification process using a simple keyword search. 
Based on user's keyword queries, the iCrawl Wizard suggests seed URLs obtained 
from Web search engines and Social Media. Additionally, it uses information extraction 
tools to suggest representative keywords and entities for the semantic crawl specification.
This way the iCrawl Wizard opens the focused crawling technology to the non-expert users 
and enables them to easily specify their crawl intention.

\begin{figure}[t]
	\centering
	\includegraphics[width=0.9\textwidth]{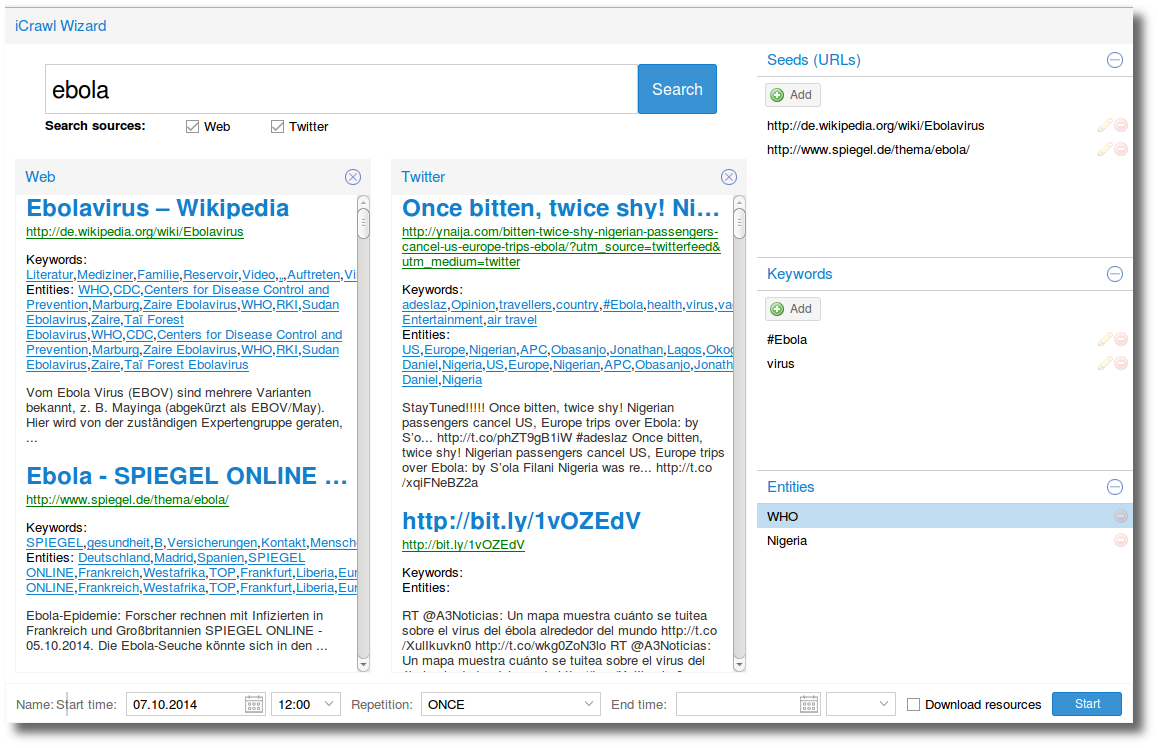}
	\caption{The iCrawl Wizard UI}
	\label{fig:crawl_spec}
\end{figure}

\section{Wizard User Interface}
\label{sec:ui}
The user interface of the \emph{iCrawl Wizard} is presented in Fig.~\ref{fig:crawl_spec}. The user of the 
iCrawl Wizard
starts by entering keywords in the search field at the top of the user interface. 
In this example, a user creating a crawl 
about the ebola outbreak enters the keyword ``ebola'' and presses the search button.
In response to this query, the system provides Web search results along 
with the results from the Twitter API.
The Web search results allow the user to find highly relevant Web pages,
while the Twitter search results provide the most recent pages about the topic.
The latter is especially important when creating a collection about a current event.

\begin{figure}[t]
  \centering
  \includegraphics[width=.9\textwidth]{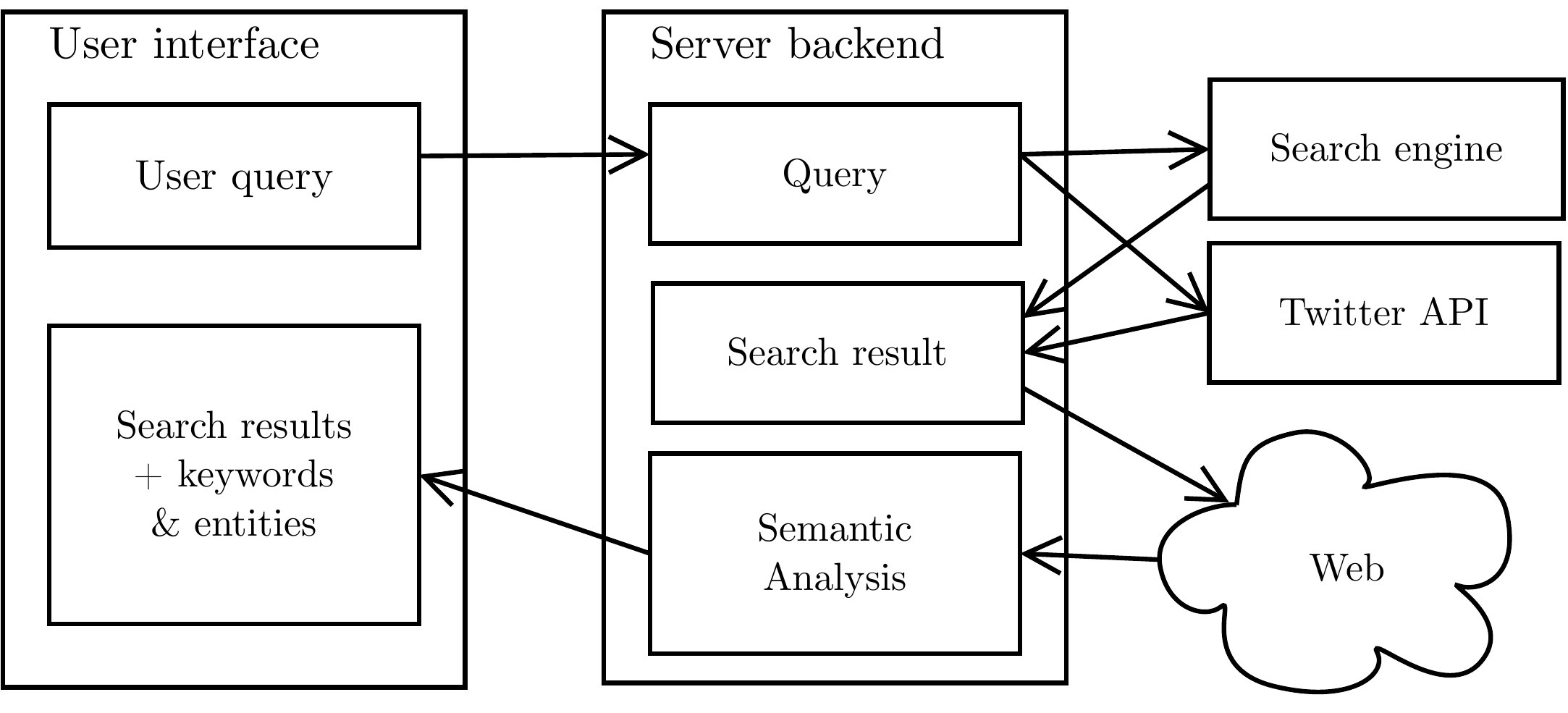}
  \caption{Architecture of the iCrawl Wizard}
  \label{fig:architecture}
\end{figure}

The search results are extended by keywords and entities describing the pages in more detail.
In this example, the first Web search result is the German Wikipedia article on Ebola, 
from which the entities such as World Health Organization (WHO) and Robert Koch Institute (RKI)
are extracted. 

Seed URLs, keywords and entities can be selected and added to the crawl specification by clicking on them.
The crawl specification constructed so far is visible on the right hand side of the interface, allowing the user to inspect and modify the list of selected seed URLs, keywords and entities. 
The user can also click on the search results to examine the Web pages before adding them 
to the crawl specification.
Items found outside the Wizard can be added to the crawl specification manually 
by clicking the corresponding ``add'' buttons.
If necessary, the user can re-formulate the search query and thus construct 
crawl specification incrementally in several interaction steps. 
Finally, further crawl parameters such as start time and duration can be specified at the bottom of the interface.

\section{Architecture}
\label{sec:architecture}

The user interface presented in Section~\ref{sec:ui} is implemented as a Web application supported by a server-side component.
The architecture is shown in Fig.~\ref{fig:architecture}.

When the user enters a keyword query, this query is sent to the server, which in term forwards the query to the search APIs of Web search engines (e.g.\@ Bing) and Social Media APIs (e.g.\@ Twitter).
The Web search APIs return a ranked list of \textit{(URL, title, description)} triples that can be processed further in this form.
The Social Media APIs return a collection of posts, so we extract the links contained in the posts and order them by their frequency.
In the case of Twitter, we also extract \emph{hashtags} (e.g.\@ ``\#ukraine'') as proposed keywords.
The text of the posts is used as a description of the extracted links.

The descriptions gained this way provide relatively few information.
Therefore we download the pages from the Web and use information extraction tools such as the Stanford Named Entity Recognizer\footnote{\url{http://nlp.stanford.edu/software/CRF-NER.shtml}} and the TextRank algorithm~\cite{MihalceaT04} to extract key terms and entities from the Web page text.
These are used to augment the results presented to the user.

All user actions such as issued queries as well as added and removed items are logged into a database.
This enables the creation of a comprehensive \emph{crawl description} to allow better sharing 
and re-use of the created document collection.

\section{Demonstration Overview}
The aim of the iCrawl Wizard is to assist users in defining a
crawl specification for a topic of interest by starting from simple keywords. 
In our demonstration we will show how the
iCrawl Wizard works and how users can use it to obtain the desired 
crawl specification without any prior knowledge about Web crawling.

During the demonstration, our audience can try the iCrawl Wizard interface.
To highlight the advantages of our approach, we will ask
our audience to perform crawl specification using the iCrawl Wizard 
as well as to suggest the seed URLs, terms and entities for the crawl specification manually. 
Through the comparison, the audience can get some hand-on experience about 
dataset creation problems on the Web. 
We will make the iCrawl Wizard available as open source software after the conference.

\small
\section*{Acknowledgments}
The authors would like to thank Bohdan Tkachenko for supporting the
implementation of the user interface.
This work was partially funded by the ERC under ALEXANDRIA (ERC 339233), 
and the COST Action IC1302 (KEYSTONE).

\small

\bibliographystyle{plain}
\bibliography{dl}

\end{document}